\documentclass{article}
\usepackage{graphicx} 
\usepackage[margin=1in]{geometry}

\usepackage{tikz}
\usetikzlibrary{positioning,calc}

\title{Why Speech Deepfake Detectors Won’t Generalize:\\ The Limits of Detection in an Open World}
\author{Visar Berisha, Prad Kadambi, Isabella Lenz}
\date{September 2025}

\begin{document}

\maketitle

\begin{abstract}
Speech deepfake detectors are often evaluated on clean, benchmark-style conditions, but deployment occurs in an open world of shifting devices, sampling rates, codecs, environments, and attack families. This creates a ``coverage debt" for AI-based detectors: every new condition multiplies with existing ones, producing data blind spots that grow faster than data can be collected. Because attackers can target these uncovered regions, worst-case performance (not average benchmark scores) determines security. To demonstrate the impact of the coverage debt problem, we analyze results from a recent cross-testing framework. Grouping performance by bona fide domain and spoof release year, two patterns emerge: newer synthesizers erase the legacy artifacts detectors rely on, and conversational speech domains (teleconferencing, interviews, social media) are consistently the hardest to secure. These findings show that detection alone should not be relied upon for high-stakes decisions. Detectors should be treated as auxiliary signals within layered defenses that include provenance, personhood credentials, and policy safeguards.
\end{abstract}
\section{Introduction}

Speech deepfakes, synthetic or manipulated speech that mimics a target speaker, are advancing rapidly in fidelity, fluency, and prosodic realism. In response, speech deepfake detectors (SDD \cite{li2025survey})  are now central to benchmarks \cite{wang2024asvspoof, yamagishi2021asvspoof, todisco2019asvspoof}, product roadmaps, and policy proposals \cite{leslie2024future}. The deployment reality, however, is not a frozen benchmark but an open, moving target: devices and microphones, acoustic environments, channels and codecs, platform post-processing (gain control, noise removal, normalization, etc.), speaking styles, emotions, languages, dialects and accents, user demographics, and attacker adaptations all vary—and they co-vary \cite{li2025survey}. Recent work has documented that systems trained on benchmark datasets suffer large performance drops when evaluated on real-world ``in-the-wild" speech, a sign that benchmark results may overestimate robustness \cite{muller2022does}. Achieving ``coverage" by training on everything is not merely expensive; it scales multiplicatively across factors and is therefore combinatorially prohibitive. If we denote the number of distinct settings along each axis by $|D|$ (devices), $|E|$ (environments), $|C|$ (channels/codecs), $|P|$ (platform transforms), $|S|$ (speaking styles), $|L|$ (languages), $|A|$ (accents/dialects), and $|T|$ (attack/tool families), then even a modest per-cell sample requirement $n$ implies a data need on the order of
\[
n \times |D|\times|E|\times|C|\times|P|\times|S|\times|L|\times|A|\times|T| \times \cdots
\]
With a fixed set of axes, growth over time is at least polynomial in the number of categories (as each $|{\cdot}|$ expands), and adding axes multiplies requirements outright. 

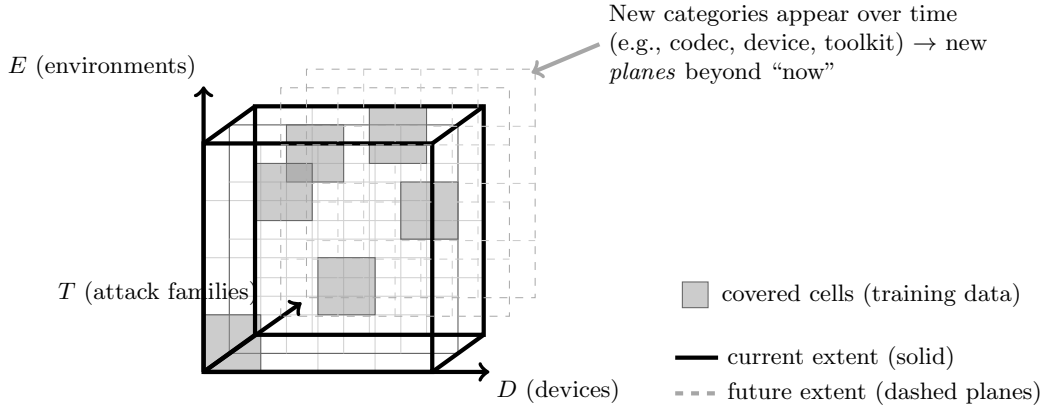
\begin{figure}[t]
\centering
\begin{tikzpicture}[scale=0.95, every node/.style={font=\small}]
  \def\N{4}
  \def\s{0.8}
  \def\dx{0.36}
  \def\dy{0.26}
  \def\Zcur{3}
  \def\Zfut{2}

  \pgfmathtruncatemacro{\Nmone}{\N-1}
  \pgfmathtruncatemacro{\Zcurmone}{\Zcur-1}
  \pgfmathtruncatemacro{\Zmax}{\Zcur+\Zfut-1}

  \pgfmathsetmacro{\FrontW}{\N*\s}
  \pgfmathsetmacro{\FrontH}{\N*\s}
  \pgfmathsetmacro{\BackShiftX}{\Zcurmone*\dx}
  \pgfmathsetmacro{\BackShiftY}{\Zcurmone*\dy}
  \pgfmathsetmacro{\BackW}{\FrontW+\BackShiftX}
  \pgfmathsetmacro{\BackH}{\FrontH+\BackShiftY}
  \pgfmathsetmacro{\FarShiftX}{\Zmax*\dx}
  \pgfmathsetmacro{\FarShiftY}{\Zmax*\dy}
  \pgfmathsetmacro{\FutCornerX}{\FrontW+\FarShiftX}
  \pgfmathsetmacro{\FutCornerY}{\FrontH+\FarShiftY}

  \foreach \z in {0,...,\Zcurmone} {
    \pgfmathsetmacro{\Xshift}{\z*\dx}
    \pgfmathsetmacro{\Yshift}{\z*\dy}
    \draw[black!65] (\Xshift,\Yshift) rectangle ++(\FrontW,\FrontH);
    \foreach \i in {1,...,\Nmone} {
      \pgfmathsetmacro{\gx}{\Xshift+\i*\s}
      \pgfmathsetmacro{\gy}{\Yshift+\i*\s}
      \draw[gray!35] (\gx,\Yshift) -- ++(0,\FrontH);
      \draw[gray!35] (\Xshift,\gy) -- ++(\FrontW,0);
    }
  }

  \foreach \x/\y/\z in {0/0/0, 2/1/0, 1/3/1, 3/2/1, 0/2/2, 2/3/2} {
    \pgfmathsetmacro{\X}{\x*\s + \z*\dx}
    \pgfmathsetmacro{\Y}{\y*\s + \z*\dy}
    \fill[black!40, fill opacity=0.5] (\X,\Y) rectangle ++(\s,\s);
    \draw[black!60] (\X,\Y) rectangle ++(\s,\s);
  }

  \draw[ultra thick] (0,0) rectangle (\FrontW,\FrontH);
  \draw[ultra thick] (\BackShiftX,\BackShiftY) rectangle (\BackW,\BackH);
  \draw[ultra thick] (0,0) -- (\BackShiftX,\BackShiftY);
  \draw[ultra thick] (\FrontW,0) -- (\BackW,\BackShiftY);
  \draw[ultra thick] (0,\FrontH) -- (\BackShiftX,\BackH);
  \draw[ultra thick] (\FrontW,\FrontH) -- (\BackW,\BackH);

  \draw[->,ultra thick] (0,0) -- (\FrontW+0.8,0) node[below right=-2pt] {$D$ (devices)};
  \draw[->,ultra thick] (0,0) -- (0,\FrontH+.8) node[above left=-2pt] {$E$ (environments)};
  \pgfmathsetmacro{\TendX}{\BackShiftX+0.65}
  \pgfmathsetmacro{\TendY}{\BackShiftY+0.45}
  \draw[->,ultra thick] (0,0) -- (\TendX,\TendY);
  \node[anchor=east] at (+0.9,\FrontH-2.1) {$T$ (attack families)};

  \foreach \k in {0,...,\numexpr\Zfut-1} {
    \pgfmathtruncatemacro{\z}{\Zcur+\k}
    \pgfmathsetmacro{\Xshift}{\z*\dx}
    \pgfmathsetmacro{\Yshift}{\z*\dy}
    \draw[gray!65,dashed] (\Xshift,\Yshift) rectangle ++(\FrontW,\FrontH);
    \foreach \i in {1,...,\Nmone} {
      \pgfmathsetmacro{\gx}{\Xshift+\i*\s}
      \pgfmathsetmacro{\gy}{\Yshift+\i*\s}
      \draw[gray!40,dashed] (\gx,\Yshift) -- ++(0,\FrontH);
      \draw[gray!40,dashed] (\Xshift,\gy) -- ++(\FrontW,0);
    }
  }

  \pgfmathsetmacro{\LblX}{\FutCornerX+0.9}
  \pgfmathsetmacro{\LblY}{\FutCornerY+0.35}
  \node[anchor=west,align=left,text width=5.2cm] (FUTNOTE) at (\LblX,\LblY)
    {New categories appear over time\\(e.g., codec, device, toolkit) $\rightarrow$ new \emph{planes} beyond ``now''};
  \draw[->,ultra thick,gray!70] (FUTNOTE.west) -- (\FutCornerX-0.05,\FutCornerY-0.05);

  \def\LegendX{\FrontW+3.5}
  \fill[black!40, fill opacity=0.5] (\LegendX, 0.9) rectangle ++(0.36,0.36);
  \draw[black!60] (\LegendX, 0.9) rectangle ++(0.36,0.36);
  \node[anchor=west] at (\LegendX+0.42, 1.08) {covered cells (training data)};

  \draw[ultra thick] (\LegendX-0.1, 0.2) -- ++(0.6,0);
  \node[anchor=west] at (\LegendX+0.5, 0.2) {current extent (solid)};

  \draw[dashed,ultra thick,gray!70] (\LegendX-0.1, -0.3) -- ++(0.6,0);
  \node[anchor=west] at (\LegendX+0.5, -0.3) {future extent (dashed planes)};
\end{tikzpicture}
\caption{Combinatorial coverage problem: The cube represents a combinations of factors (e.g., devices, environments, attack families). Solid planes mark the current space of conditions (a 3-factor example), with shaded cells indicating the limited portions actually covered by training data. Unshaded cells represent uncovered regions that can serve as attack surfaces for adversaries. Dashed planes show how new categories that emerge over time expand the space beyond today’s extent, multiplying the number of uncovered combinations. This growing gap between required and available coverage constitutes the ``coverage debt."} \label{fig:covdebt}
\end{figure}

A natural response is to ``just add more data" to cover cells in $(D,E,C,P,S,L,A,T)$. Collecting more data helps, but it cannot close the gap. Figure \ref{fig:covdebt} illustrates why for a simplified set of axes: growth is multiplicative while collection is additive. Each new codec, device, or attack family expands into a full cross-product with all other axes, so the number of required cells grows far faster than data can be gathered. The shaded regions in the figure show the limited portion of the current space covered by training data, while the unshaded cells represent potential attack surfaces available to adversaries. We term the accumulating gap between required coverage and available data the ``coverage debt." Second, many cells are intrinsically hard to fill as they are rare in the wild, privacy-restricted, or otherwise inaccessible (e.g., low-resource dialects, secure workflows) \cite{almutairi2022review}. Third, new cells emerge only after deployment, when novel codecs, platforms, or synthesizer toolkits appear, making the distribution fundamentally non-stationary. Fourth, risk is adversarially weighted: attackers deliberately seek the uncovered or weakest cells, so worst-case performance, not average coverage, determines security. Unlike other applications in speech technology (speech recognition, speaker ID, etc) that can generalize across tasks by leveraging shared structure in human speech, deepfake detection cannot amortize its coverage debt in the same way. Residual differences between real and synthetic audio are inconsistent across coverage domains. Because SDD relies on these residual differences, there is little stable signal to transfer or share. Consequently, this debt cannot be repaid by incremental data accrual or fine-tuning over a handful of datasets; the domain is unbounded and time-varying. The key implication is that if we want to use speech deepfake detection to make decisions for real-world problems involving deepfakes, then detectors must be robust to evolving world conditions rather than optimized for fixed benchmarks. This motivates our deployment-facing question in this paper: can detectors trained on one snapshot reliably decide under different conditions?

To help answer this question, we use the evaluation strategy proposed by Kwok et~al. \cite{kwok25}. They show that the common practice of reporting a single EER on a pooled dataset, typically dominated by clean, read speech, can overstate detector performance. They advocate \emph{bona fide cross-testing}: pairing each synthesizer (one per subset) with multiple bona fide datasets and reporting EER for every pairing, then aggregating across pairings for a balanced summary.

We operationalize the moving-target problem in the equation above using this evaluation framework without attempting exhaustive coverage. We re-analyze the data from \cite{kwok25} along two deployment-salient axes: (a) bona fide \emph{domain}—a proxy for the device/environment/channel/style bundle $(D,E,C,P,S)$—and (b) spoof \emph{release year}, a proxy for the evolving attacker/tool family $T$ and non-stationarity. We summarize performance across detectors without per-pair re-training. Two patterns emerge. First, a step-change in difficulty coincides with post-2022 synthesizers: while earlier families can look tractable under a standard error metric, newer systems erase the legacy artifacts some SDD models rely on. Second, even within a fixed calendar year, variability across bona fide domains is large; conversational settings (teleconferencing, interviews, social media) are consistently harder than read or broadcast-style speech. 

The experiment highlights the deployment risk associated with deepfake detection across the open set of real-world conditions, with the worst failures concentrated in the channels adversaries are likely to exploit. We posit that for applications where reliable knowledge that a human is speaking is essential (e.g., account recovery, high-risk  actions), detector-only defenses are insufficient and other approaches to this problem are necessary.

\section{Methods}

Kwok et al. \cite{kwok25} introduced a bona fide cross-testing framework for audio deepfake detection, in which each spoof subset (one synthesizer family) is paired with multiple bona fide datasets of varying domains. Instead of reporting a single pooled Equal Error Rate (EER), they recommend computing per-pair EERs and then summarizing across pairings to expose variability across spoof families and bona fide speech types. They demonstrate their evaluation framework using three detectors based on semi-supervised learning (\emph{Wav2Vec-Conformer}, \emph{Wav2Vec-TCM}, \emph{Wav2Vec-SCL)}, nine bona fide datasets, and more than 160 spoof subsets released between 2019 and 2024.

Using this evaluation framework, we analyze the cross-testing results with a deployment focus. Specifically, we concentrate on the strongest detector (\emph{Wav2Vec-SCL}) evaluated in the original paper and compute per-pair EERs as in \cite{kwok25}, but then regroup the results by (i) application-focused domain of the bona fide set and (ii) release year of the spoof subset. 

\vspace{0.5cm}

\noindent {\bf Speech Deepfake Detector:} The highest-performing detector in \cite{kwok25} is the \emph{Wav2Vec-SCL} model, introduced by Doan et al. \cite{doan2024balance}. The detector uses a large self-supervised speech representation model (XLS-R \cite{conneau2020unsupervised}) as its feature extractor, followed by a lightweight classifier consisting of three fully connected layers with ReLU activations and a final dense layer with log-softmax outputs for the two classes (bona fide vs. spoof). To improve generalization across spoofing methods and recording conditions, the model is trained with: (i) balanced sampling to ensure equal representation of real and spoofed speech in mini-batches, (ii) multiple augmentation methods (e.g., additive noise, reverberation, codec distortion) applied to both real and fake speech, and (iii) inclusion of re-synthesized samples, in which bona fide recordings are vocoded with diverse neural vocoders to generate harder negatives. For full details we refer the reader to \cite{doan2024balance}.
\vspace{0.5cm}

\noindent {\bf Performance Metric:} The performance metric for the SDD is the equal error rate (EER). The EER is the operating point where the false positive rate equals the false negative rate (or is closest, in the discrete case). At this threshold both types of errors occur at the same rate, and that common value is reported as the EER. This error metric is often reported in SDD benchmarks \cite{li2025survey, wang2024asvspoof, yamagishi2021asvspoof, todisco2019asvspoof}.
\vspace{0.5cm}

\noindent {\bf Evaluation data:} We evaluate \emph{Wav2Vec-SCL} in pairwise fashion across multiple bona fide vs. spoof subsets. We use the nine bona fide datasets from \cite{kwok25} and map the nine bona fide subsets to deployment-salient domains as shown in Table \ref{tab:bf_domain_map}. As in \cite{kwok25}, we compare each bona fide dataset against more than 160 spoofed speech models, drawn from a broad range of public corpora spanning 2019–2024. These include the official ASVspoof benchmarks (ASVspoof 2019 LA, ASVspoof 2021 DF), as well as multiple community and research releases such as FakeAVCeleb \cite{khalid2021fakeavceleb}, EmoFake-EN \cite{zhao2024emofake}, AV-Deepfake-1M \cite{cai2024av}, CodecFake \cite{xie2025codecfake}, MLAAD-v3-EN \cite{muller2024mlaad}, and LlamaPartialSpoof \cite{luong2025llamapartialspoof}. Each spoof subset corresponds to a distinct synthesizer family (e.g., neural TTS systems, voice conversion models, vocoders, codec-based manipulations), so the $>160$ total reflects the diversity of generation pipelines and parameterizations represented across these corpora. In total, the spoof sets span both older systems (pre-2022) and newer models (2022–2024), providing coverage of multiple generations of synthesis technology.

\begin{table}
\centering
\caption{Bona fide subsets and their mapped domains (after \cite{kwok25}).}
\label{tab:bf_domain_map}
\begin{tabular}{ll}
\hline
\textbf{Dataset} & \textbf{Mapped domain} \\
\hline
AMI IHM (meeting \cite{carletta2005ami} & Teleconferencing / meetings \\
AMI SDM (meeting) \cite{carletta2005ami} & Teleconferencing / meetings \\
LibriSpeech test-clean \cite{panayotov2015librispeech} & Audiobooks  \\
LibriSpeech test-other \cite{panayotov2015librispeech} & Audiobooks  \\
VCTK 0.92 \cite{yamagishi2019cstr} & News / broadcast-style read speech \\
FakeAVCeleb-v1.2 \cite{khalid2021fakeavceleb} & Interviews / podcasts \\
In-The-Wild \cite{muller2022does} & Social media \\
EmoFake-EN \cite{zhao2024emofake} & Emotional / acted speech \\
AV-Deepfake-1M \cite{cai2024av} & Interviews / podcasts \\
\hline
\end{tabular}
\end{table}

\vspace{0.5cm}
\noindent {\bf Statistical Analysis:} For each domain $d$ (set of bona fide IDs) and spoof release year $y$, we aggregate the pairwise EERs across all synthesizer subsets released in year $y$, after first averaging across bona fide sets within domain $d$. Concretely, letting $\mathcal{K}(d)$ be the bona fide datasets in domain $d$ and $\mathcal{M}(y)$ the spoof subsets released in year $y$,
\[
\overline{\mathrm{EER}}(d,y)=
\frac{1}{\,|\mathcal{M}(y)|} \sum_{m\in\mathcal{M}(y)}
\left(\frac{1}{|\mathcal{K}(d)|}\sum_{k\in\mathcal{K}(d)} \mathrm{EER}_{k,m}\right).
\]
We report \emph{mean} $\pm$ \emph{SEM}, where the SEM at $(d,y)$ is the standard error of the set
$\{\, \frac{1}{|\mathcal{K}(d)|}\sum_{k} \mathrm{EER}^{(\ell)}_{k,m}\,\}_{m\in\mathcal{M}(y)}$. All EERs are shown as percentages.

\section{Results}

\begin{figure}
    \centering
    \includegraphics[width=0.6\linewidth]{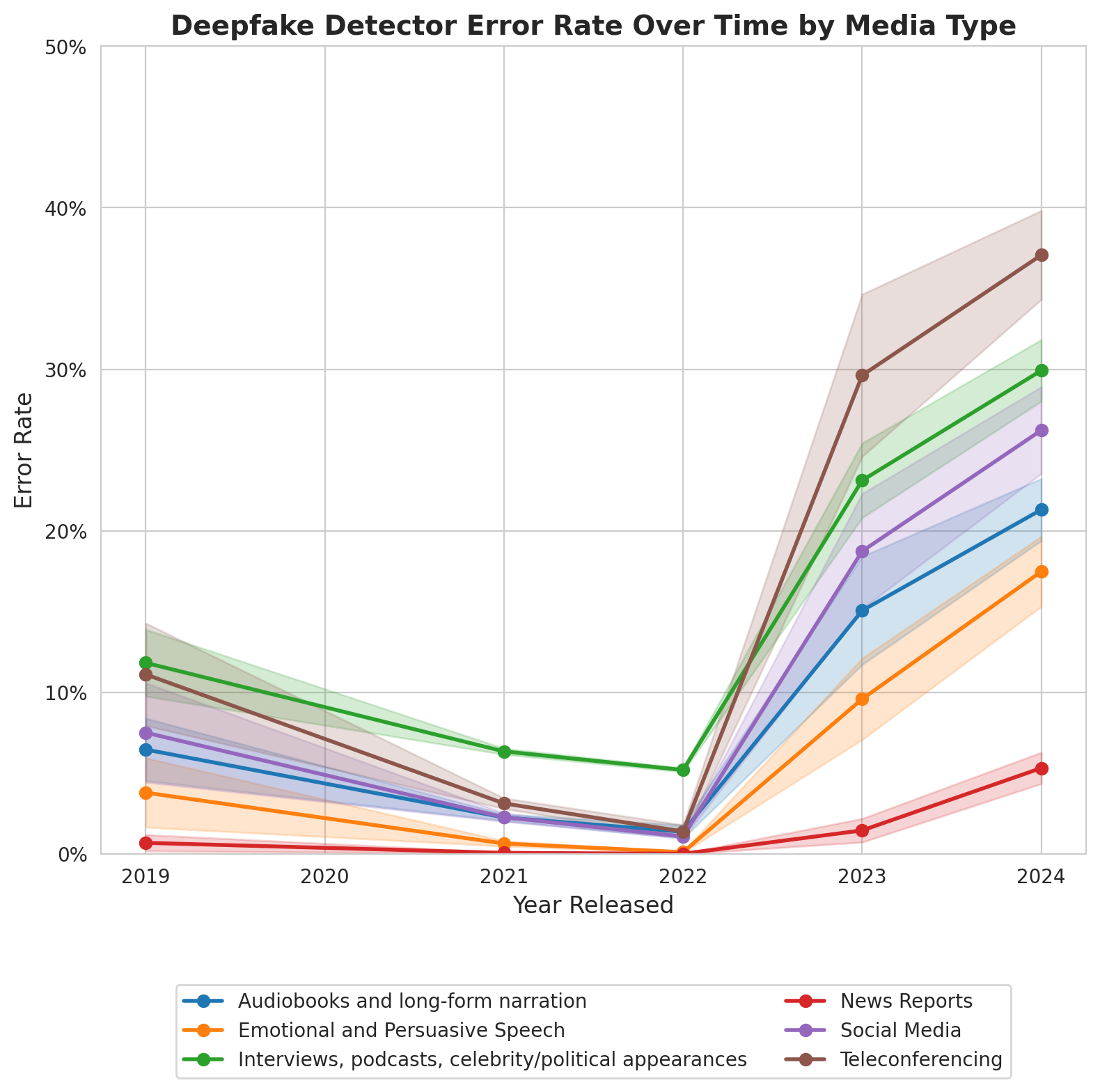}
    \caption{Detection performance summarized by bona fide domain and spoof release year. The line shows the mean Equal Error Rate (EER) and the standard error of measure.}
    \label{fig:domain_trends}
\end{figure}

Figure~\ref{fig:domain_trends} summarizes detector performance by domain (colors) and spoof release year ($x$-axis). Three consistent patterns emerge. First, errors are comparatively low for legacy spoof families (2019, 2021) and remain modest for 2022, but increase sharply for 2024 across nearly all domains. Second, the rise is domain-dependent: conversational settings—teleconferencing/meetings, interviews/podcasts, and social media - exhibit the largest degradation, whereas read speech remains the most tractable. Third, uncertainty bands (SEM) widen in 2024, reflecting heterogeneity across newer synthesizer families and sensitivity to detector choice; nevertheless, the trend (higher mean EER in 2024) is consistent across all three detectors. 

\section{Discussion}

Using bona fide cross-testing by domain and spoof release year to reflect open-world conditions, we find that detectors that appear reliable on legacy synthesizers do not generalize to newer spoof families, with the steepest degradations in conversational settings and widening uncertainty across spoofing models (Figure~\ref{fig:domain_trends}). The pronounced errors on 2024 spoofs exemplify the coverage debt problem: incremental additions of training data are insufficient against the multiplicative growth of new synthesizer families, speaking styles, and deployment conditions. That read speech remains comparatively tractable underscores that performance on benchmark conditions with homogeneous data is not a reliable proxy for robustness in dynamic, real-world scenarios. The widening uncertainty bands further highlight that heterogeneity in new attacks and detector sensitivities compounds this debt, leaving unshaded ``cells" in Figure~\ref{fig:covdebt} that adversaries can exploit.

\begin{figure}[ht]
    \centering
    \includegraphics[width=0.6\linewidth]{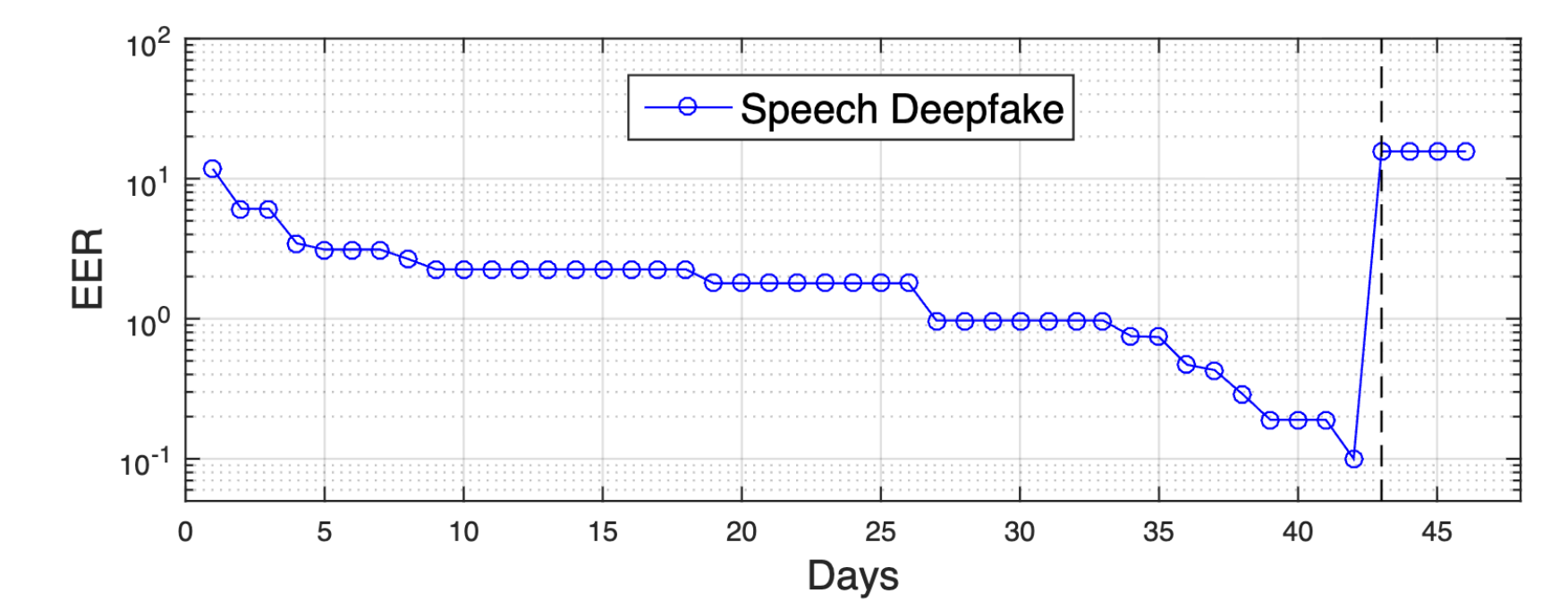}
    \caption{EER of the best-performing system in the ASVspoof 2021 deepfake challenge. Errors fall during training (days 1–42) but spike on day 43 when the evaluation set introduces new spoofing attacks, codecs, and conditions, highlighting the generalization gap to unseen cells in the factor space. From \cite{yamagishi2021asvspoof}.}
    \label{fig:EERasvspooof21}
\end{figure}

Further demonstration of this generalization gap is provided by the ASVSpoof21 deepfake (DF) detection challenge. Figure \ref{fig:EERasvspooof21} shows the EER of the best-performing model across the model optimization phase (days 1–42) on the held-out set and during evaluation (day 43 onward). While error steadily declines during the model optimization phase, the evaluation set produces a sudden jump in EER to levels even higher than those observed on day 0. This discontinuity arises because the evaluation set introduces new spoofing attacks, compression methods, and data conditions absent from the training phase. In terms of our framing, the evaluation set contains data from new cells (and possibly planes) in the factor space (Figure\ref{fig:covdebt}), leaving blind spots where the model fails. 

Coverage blind spots may explain many of the vulnerabilities that detectors are likely to face, but they are only part of the challenge in speech deepfake detection. The gaps in Fig. \ref{fig:covdebt} reflect {\em epistemic uncertainty} — blind spots in the factor space where detectors have not been trained (e.g., unseen devices, environments, spoofing families) \cite{hullermeier2021aleatoric}. Our analysis primarily illustrates this form of uncertainty: coverage debt that arises whenever new conditions appear. But epistemic uncertainty is not the only obstacle. {\em Aleatoric uncertainty} stems from irreducible variability \cite{hullermeier2021aleatoric}: as synthesis quality improves and real-world channels add noise, compression, or sampling artifacts, bona fide and spoofed speech increasingly overlap. This overlap grows as newer systems erase legacy artifacts, leaving detectors unable to distinguish real from synthetic when natural variability in prosody, noise, or codecs mimics spoofing cues. Unlike epistemic uncertainty, which can in principle be reduced by expanding training coverage, aleatoric uncertainty is irreducible. It guarantees a non-zero error floor for any detector, ensuring that deepfake detection alone is likely insufficient for reliable real-world deployment, particularly as synthetic data grows closer to real.

The challenges of coverage and adaptation are not unique to deepfake detection. In cybersecurity, cat-and-mouse dynamics are the norm: spam filters, malware detection, and intrusion prevention systems all operate under incomplete coverage and adversarial pressure, yet their attack surfaces are narrower and better bounded. Spam filters contend with a limited set of textual and structural cues that can be updated quickly. Malware detectors, though faced with mutating binaries, are constrained by instruction sets and operating system behaviors, and can be supplemented with sandboxing and behavior analysis. Intrusion detection systems rely on deviations from well-specified protocols and traffic baselines. In speech technology, systems such as ASR, speaker recognition, and voice biometrics succeed by exploiting stable structure in human speech—phonetics, syntax, and speaker-specific traits—that persists across channels and conditions and supports generalization. Deepfake detection, by contrast, faces both an unbounded and adversarial factor space—devices, codecs, languages, speaking styles, spoofing families—and a reliance on residual differences between real and synthetic speech that shrink as synthesis improves. This combination makes coverage debt unusually severe: blind spots expand faster than data can be collected, and even perfect coverage cannot prevent a persistent error floor once bona fide and spoof distributions overlap.

Because these uncertainties are structural rather than model-specific, we expect our empirical results with \emph{Wav2Vec-SCL} to generalize. \emph{Wav2Vec-SCL} was the best-performing system in Kwok et al.’s benchmark \cite{kwok25}, consistently outperforming the other detectors they tested. If this detector shows systematic degradation across domains and years, weaker detectors are unlikely to fare better. This same pattern is evident in benchmark evaluations such as the ASVspoof challenges, where systems that perform well on development sets often collapse under novel spoof families, codecs, or domains. In related work, the authors in \cite{muller2022does} find that many architectures that perform well in lab settings suffer dramatic drops in performance when tested on ``in-the-wild" data (real-world recordings of celebrities and public figures) demonstrating that the generalization performance of benchmark results is overstated. Moreover, the underlying challenges - coverage debt (epistemic uncertainty) and irreducible overlap (aleatoric uncertainty) - do not depend on a specific architecture: any detector trained on finite data snapshots will face the same coverage explosion as conditions diversify, and any system that relies on statistical differences between real and synthetic speech will encounter overlap as generative models improve and channel effects blur class boundaries.

\subsection{Implications for Deployment}

Our experiments provide insight into how speech deepfake detectors should be deployed. First, detectors should not be treated as primary gates for high-stakes decisions (e.g., financial transactions, authentication, or public safety messaging). Instead, they are best used as \emph{monitoring or triage tools}: to flag suspicious inputs, trigger secondary checks, or provide risk scores that are combined with other signals. Operating thresholds must be domain-specific and adaptive over time, rather than fixed globally, and regular auditing and recalibration are essential to manage drift.

Second, adversarial adaptation amplifies the challenge. As benchmarks and models become public, attackers can tune their synthesis pipelines to remove  the cues detectors rely on, accelerating the disappearance of detectable artifacts. This dynamic ensures that any detector optimized for a fixed snapshot will eventually be outpaced by new families of attacks, underscoring the need for continuous monitoring, periodic re-evaluation, and integration with complementary defenses rather than reliance on a static model.

Third, while expanding training data may reduce epistemic uncertainty in the short term, it cannot solve the problem at scale. The cross-product of devices, codecs, environments, speaking styles, languages, and spoofing families grows faster than data can be collected, and many high-risk domains (such as banking calls or teleconferences) are legally or practically inaccessible. Coverage debt is therefore a permanent feature of the problem, not a temporary gap.

Finally, robust mitigation requires a layered defense. Provenance mechanisms such as capture-time attestation, personhood credentials, challenge–response protocols, and multi-channel corroboration can add information that attackers cannot easily spoof. Detector scores retain value in this ecosystem as one signal among many, but they must be integrated with policy and UX design that set realistic expectations. 

\section{Conclusion}

Our experiment demonstrates that speech deepfake detectors that perform well on legacy benchmarks do not generalize to newer spoofing families or conversational domains. The pronounced errors on 2024 synthesizers exemplify coverage debt: incremental data additions cannot keep pace with the combinatorial growth of devices, codecs, styles, and attack families. While coverage gaps reflect epistemic uncertainty and may be narrowed with additional data, aleatoric uncertainty—irreducible overlap between bona fide and synthetic speech under realistic conditions—guarantees a non-zero error floor for any detector. Because these uncertainties are structural, not architectural, the limitations we observe with \emph{Wav2Vec-SCL} extend to detectors more broadly. This implies that detector-only defenses are insufficient for deployment in open-world conditions. 

\bibliographystyle{IEEEtran}
\bibliography{references}

\end{document}